\documentclass[prl,twocolumn,a4paper,nofootinbib,showpacs]{revtex4-1}

\usepackage{hyperref}
\usepackage{ulem}
\usepackage{graphicx}
\usepackage{bm}
\usepackage{amsmath}
\usepackage{amssymb}
\usepackage{amsthm}
\usepackage{color}
\usepackage[nice]{nicefrac}

\theoremstyle{plain}

\renewcommand{\O}{\mathcal{O}}
\newcommand{\ket}[1]{|#1\rangle}
\newcommand{\bra}[1]{\langle #1 |}

\begin{document}

\title{Observers can always generate nonlocal correlations without aligning measurements by covering all their bases}

\author{Joel J. \surname{Wallman}}
\affiliation{School of Physics, The University of Sydney,
Sydney, New South Wales 2006, Australia}
\author{Stephen D. \surname{Bartlett}}
\affiliation{School of Physics, The University of Sydney,
Sydney, New South Wales 2006, Australia}

\date{6 February 2012}

\begin{abstract}
Quantum theory allows for correlations between the outcomes of distant measurements that are inconsistent with any locally causal model, as demonstrated by the violation of a Bell inequality. Typical demonstrations of these correlations require careful alignment between the measurements, which requires distant parties to share a reference frame. Here, we prove, following a numerical observation by Shadbolt et al., that if two parties share a Bell state and each party preforms measurements along three perpendicular directions on the Bloch sphere, then the parties will always violate a Bell inequality. Furthermore, we prove that this probability is highly robust against local depolarizing noise, in that small levels of noise only decrease the probability of violating a Bell inequality by a small amount. We also show that generalizing to $N$ parties can increase the robustness against noise. These results improve on previous ones that only allowed a high probability of violating a Bell inequality for large numbers of parties.
\end{abstract}


\maketitle

One of the most fascinating and useful features of quantum theory is that the correlations between the outcomes of spatially separated measurements can be nonlocal, i.e., inconsistent with any locally causal model~\cite{bell2004,barret2005a}. Typically, to obtain nonlocal correlations experimentally, great care is taken to choose measurements that give the strongest nonlocal correlations possible, which requires distant parties to share a reference frame~\cite{aspect2004,bartlett2007}. While there are proposals for violating a Bell inequality without the need for a prior shared reference frame~\cite{bartlett2007,cabello2003}, these proposals add substantial complexity to the simple form of a standard Bell test.

Distant parties could attempt to violate a Bell inequality without aligning reference frames by performing measurements in random directions~\cite{ashhab2007,ycliang2010,Wallman2011}, and recent results prove that such a method can demonstrate a violation with some nonzero probability.  Specifically, for $N$ spatially-separated parties who share a Greenberger-Horne-Zeilinger (GHZ) state~\cite{GHZ}, almost all choices of two measurements at each site lead to nonlocal correlations between measurement outcomes if the number of parties $N$ is large~\cite{ycliang2010}. Therefore distant parties that do not share a reference frame can randomly choose measurements that violate some Bell inequality with a probability that approaches 1 as $N$ increases. If the parties also share a single direction on the Bloch sphere (as can be the case in, e.g., photon polarization encodings~\cite{bartlett2007}), then they can always violate one of two Bell inequalities by an amount that is exponential in $N$~\cite{Wallman2011}.

These results are the weakest for the scenario most relevant to experiments, namely, the bipartite $N=2$ case: the probability of violating a Bell inequality by choosing two mutually unbiased measurements randomly in the bipartite case is $\sim42\%$~\cite{ycliang2010}. In this paper, we show that if the two parties each choose three measurements corresponding to the $x$, $y$ and $z$ components of their local Cartesian reference frame (hereafter referred to as a \textit{triad of measurements}), then they will always violate a Bell inequality. That this scheme always results in a violation of a Bell inequality was communicated to us by the authors of \cite{Shadbolt2011} as a conjecture. In this paper, we prove this conjecture; we note that Ref.~\cite{Shadbolt2011} presents an independent proof of the same. We also prove that this form of a Bell test is robust against noise, in that small levels of noise only slightly decrease the probability of violating a Bell inequality.

For the multipartite case, we numerically estimate the probability of $N$ parties who share an $N$-partite GHZ state violating a Bell inequality as a function of the level of local depolarizing noise when the $N$ parties each choose a triad of measurements. In the absence of noise, we find that the parties will always violate a Bell inequality (except for $N=3$, where the numerical probability is ${\sim 99.99\%}$) and the robustness to noise increases with $N$.

An intuitive way of understanding the success rate of this scheme is as follows. When the parties each choose a triad of measurements, one of each parties' three measurements must necessarily be within an angle $\frac{\pi}{3}$ of the $z$-axis of the reference frame in which the entangled state was created. Although the parties do not know which of their measurement directions are closest to the $z$-axis, by choosing a triad of measurements they have covered all possibilities and so can simply test each possibility using the method of Ref.~\cite{Wallman2011} to always obtain nonlocal correlations. It has previously been observed that parties that do not share a reference frame can obtain nonlocal correlations by trying all possible combinations of local measurement directions~\cite{ashhab2007}. While this is evidently true, it is also experimentally infeasible. However, our results show that the parties only need to try combinations of a finite (and relatively small) number of measurements at each site in order to always obtain nonlocal correlations.

\textit{The scenario.}---A verifier prepares many copies of an $N$ qubit state $\rho$ and distributes one qubit to each of $N$ parties. The $n$-th party chooses a triad of measurements, which can be written as $\O_n^{s_n} ={\Omega}_n^{s_n}\cdot\vec{\sigma}$, where $\vec{\sigma}= \left(\sigma_x, \sigma_y, \sigma_z \right)$ is the vector of Pauli matrices (relative to the reference frame in which $\rho$ was created) and $\{\Omega_n^0,\Omega_n^1,\Omega_n^2\}$ are orthonormal vectors in the Bloch sphere. The qubits are distributed over channels that introduce a level $\gamma\in[0,1]$ of local depolarizing noise, where $\gamma=1$ corresponds to no noise. Regardless of its physical origin, local depolarizing noise can be modeled by reducing the visibility of the measurements at each site as $\gamma \Omega$~\cite{Laskowski2010}. We do not consider colored noise, such as local dephasing noise, as such noise models could allow the parties to establish some common direction. This in turn would allow the parties to use the method in Ref.~\cite{Wallman2011} to obtain a greater violation of a Bell inequality with a smaller number of measurement settings and inequalities.

For each copy of $\rho$, each party randomly chooses and performs one of their three measurements on their qubit. The parties then send the verifier a list of the measurement choice, $s_n\in\mathbb{Z}_3$, and corresponding outcome, $o_n^{s_n}\in\mathbb{Z}_2$, for each copy of $\rho$. The verifier uses the lists to determine if the measurement outcomes are inconsistent with a locally causal model. 

In general, the verifier will need to use the full joint probability distributions $p(\vec{o}|\vec{s})$ to determine if the relation between the measurement outcomes $\vec{o}=(o_1,\ldots,o_N)$ and measurement settings $\vec{s} = (s_1,\ldots,s_N)$ is inconsistent with a locally causal model. However, for the scenario we consider, the verifier only needs to calculate the probabilities $p(a|\vec{s})$ (as relative frequencies) that the outcomes satisfy $\bigoplus_{n=1}^N o_n^{s_n}=a$ for $a=0,1$ (where addition is modulo 2). They can then determine the correlation functions
\begin{align}\label{eq:singlet_correlations}
E\left(\vec{s}\right) &= p(0|\vec{s}) - p(1|\vec{s})	\,,
\end{align}
and determine if the correlation functions are inconsistent with any locally causal model by checking if they violate some Bell inequality.

The Bell inequalities we consider are the Mermin-Ardehali-Belinskii-Klyshko (MABK) Bell inequalities~\cite{MABK}, which only depend on two measurement settings at each site. For $N=2$, the MABK Bell inequalities reduce to the famous Clauser-Horne-Shimony-Holt (CHSH)~\cite{CHSH} Bell inequalities.

To use Bell inequalities that only depend on two measurement settings at each site, the verifier can simply choose one setting $t_n$ for each site and ignore any copy of $\rho$ where the $n$-th party performed the measurement corresponding to $t_n$ for any value of $n=1,\ldots,N$. Mathematically, this can be represented by the verifier choosing integers $r_n\in\mathbb{Z}_2$ and injective functions $\tau_n:\mathbb{Z}_2\to\mathbb{Z}_3$ for $n=1,\ldots,N$. The measurement settings that the verifier checks for the $n$-th site are $s_n\in\{\tau_n(0),\tau_n(1)\}$. Denoting by $\tau(\vec{r})=(\tau_1(r_1),\ldots,\tau_N(r_N))$ the set of measurement settings corresponding to a specific choice of $\vec{r}=(r_1,\ldots,r_N)\in\mathbb{Z}_2^N$, all MABK Bell inequalities can be obtained from the inequality
\begin{align}\label{ineq:bell}
\Bigl|\sum_{\vec{r}\in\mathbb{Z}_2^N} \cos\left(\tfrac{R\pi}{4}\right)E\left(\tau(\vec{r})\right)\Bigr|
 \le2^{\tfrac{N-1}{2}}\,,
\end{align}
where $R = N + 1 - 2\sum_{k=1}^N r_n$, by varying over the $6^N$ functions $\tau$~\cite{Wallman2011}. Note that as presented, this is a form of post-selection, but it does not introduce a communication loop-hole as it can also be viewed as a form of pre-selection. For example, with each qubit, the verifier could also send an integer corresponding to a setting that the parties cannot use to measure that qubit. 

The different functions $\tau$ correspond to the different labelings of the measurement directions, $\Omega_n^{s_n}$. We can exploit these labelings to restrict the relative orientations of the triads of measurements $\{\Omega_n^0,\Omega_n^1,\Omega_n^2\}$. It is important to note that the verifier can only relabel the measurements in the following manner if they know the orientation between the parties' reference frames. Without such knowledge, the verifier would still have to test a variety of labelings in order to identify which measurements violate a Bell inequality.

With respect to the verifier's reference frame (in which the state $\rho$ is prepared), the $n$-th party's measurement directions can be written as
\begin{align}\label{eq:measurements}
\Omega_n^{0} &= x'_n \cos\chi_n + y'_n	\sin\chi_n \nonumber\\
\Omega_n^{1} &= -x'_n \sin\chi_n + y'_n \cos\chi_n \nonumber\\
\Omega_n^{2} &= \left(\sin\theta_n\cos\phi_n,\sin\theta_n\sin\phi_n,\cos\theta_n\right)	\,,
\end{align}
where 
\begin{align}
x'_n &= \left(\sin\phi_n,-\cos\phi_n,0\right)	\nonumber\\
y'_n &= \left(\cos\theta_n\cos\phi_n,\cos\theta_n\sin\phi_n,-\sin\theta_n\right)	\,,
\end{align}
$\theta_n\in[0,\pi]$ and $\phi_n,\chi_n\in[-\pi,\pi]$. For each $n$, one of the measurement directions $\Omega^i_n$ must be within an angle of $\nicefrac{\pi}{3}$ of either the $\pm z$ axis. We relabel the $n$-th party's measurements so that this direction is $\Omega_n^2$ and swap the sign of $\Omega_n^2$ if necessary (which corresponds to relabeling the measurement outcomes of the measurement $\O_n^{s_n}$), so that $\theta_n$ is in the interval $[0,\nicefrac{\pi}{3}]$. Adding multiples of $\nicefrac{\pi}{2}$ to $\chi_n$ simply permutes $\{\pm\Omega_n^0,\pm\Omega_n^1\}$, so we can also set $\chi_n$ to be in the interval $[-\nicefrac{\pi}{4},\nicefrac{\pi}{4}]$ for all $n$.

\textit{The bipartite case.}---As we now prove, two parties who share the singlet state,
\begin{align}
\ket{\Psi^-} = 2^{-\nicefrac{1}{2}}\left(\ket{0}\ket{1} - \ket{1}\ket{0}\right)	\,,\label{eq:singlet_state}
\end{align}
where $\ket{0}$ and $\ket{1}$ are the computational basis states in the verifier's reference frame, in the above scenario will always violate a Bell inequality. Note that this proof holds for any maximally entangled two-qubit state due to local equivalence, but we choose the singlet state for clarity. We also prove that this result is robust against local depolarizing noise. 

\begin{figure}[t!]
\centering
\includegraphics[width=.5\linewidth]{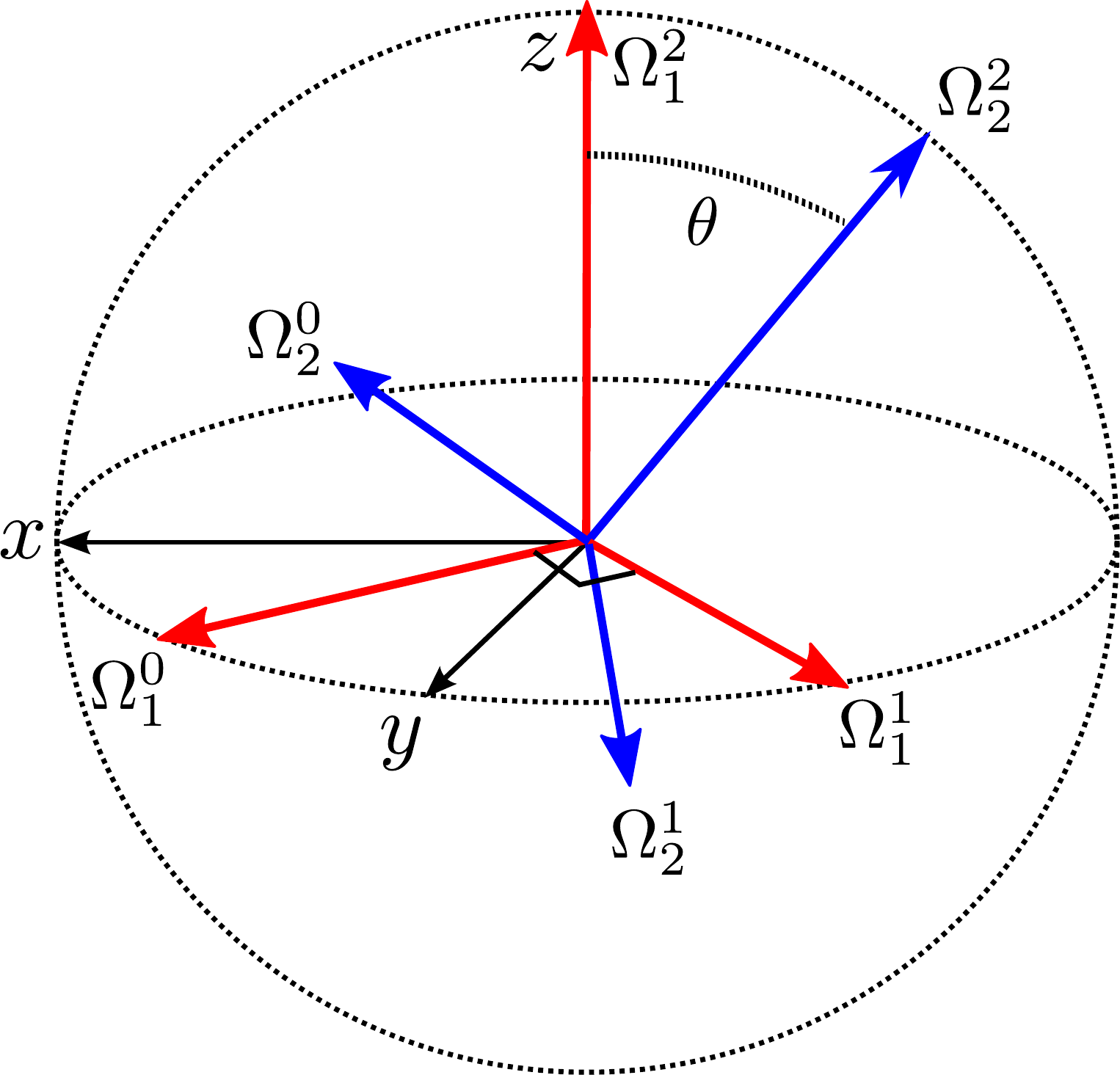}
\caption{\label{fig:restrict_theta_singlet}(Color online) Illustration of the labeling of the two parties' measurements in Eq.~\eqref{eq:singlet_alice} and \eqref{eq:singlet_bob}.}
\end{figure}

For a singlet state distributed over channels that introduce local depolarizing noise parametrized by $\gamma$, the correlation functions are
\begin{align}\label{eq:singlet_correlation}
E(s_1,s_2) &= \bra{\Psi^-}\left(\gamma\O^1_{s_1}\otimes\gamma\O^2_{s_2}\right)\ket{\Psi^-}	\nonumber\\
&= -\gamma^2\Omega^1_{s_1}\cdot\Omega^2_{s_2}	\,.
\end{align}
The singlet state is invariant under arbitrary joint rotations of the two parties' Bloch spheres, which allows us to reduce the problem to one with only three parameters. To do this, first note that we can rotate both parties' measurements so that $\Omega_1^2$ is the $z$ axis, i.e., 
\begin{align}\label{eq:singlet_alice}
\Omega_1^0 &= \left(\sin\chi_1,-\cos\chi_1,0\right)	\nonumber\\
\Omega_1^1 &= \left(\cos\chi_1,\sin\chi_1,0\right)	\nonumber\\
\Omega_1^2 &= \left(0,0,1\right)	\,,
\end{align}
where we have incorporated $\phi_1$ into $\chi_1$. We then relabel the second party's measurements so that $\Omega_2^2$ is within $\nicefrac{\pi}{3}$ of $\Omega_1^2$ (i.e., the $z$ axis). Finally, we can rotate both parties' measurements around the $z$ axis so that $\Omega_2^2$ is in the $xz$ plane, i.e., the second party's measurement directions can be written as
\begin{align}\label{eq:singlet_bob}
\Omega_2^0 &= \left(\sin\chi_2\cos\theta,-\cos\chi_2,-\sin\chi_2\sin\theta\right)	\nonumber\\
\Omega_2^1 &= \left(\cos\chi_2\cos\theta,\sin\chi_2,-\cos\chi_2\sin\theta\right)	\nonumber\\
\Omega_2^2 &= \left(\sin\theta,0,\cos\theta\right)	\,,
\end{align}
for some $\theta\in[0,\nicefrac{\pi}{3}]$ and $\chi_2\in[-\nicefrac{\pi}{4},\nicefrac{\pi}{4}]$ and the first party's measurement directions are as in Eq.~\eqref{eq:singlet_alice} for some new value of $\chi_1$ which we can set to be in the interval $[-\nicefrac{\pi}{4},\nicefrac{\pi}{4}]$ by relabeling the first party's measurements. An example of a set of measurements labeled in this manner is shown in Fig.~\ref{fig:restrict_theta_singlet}.

When the measurements are labeled in this way, the measurement statistics only need to be tested against a small number of the 36 Bell inequalities in order to demonstrate nonlocal correlations. In particular, we only need the 3 inequalities obtained from Eq.~\eqref{ineq:bell} for $\tau=(r_1,1-r_2)$, $(1-r_1,r_2)$ and $(2-r_1,1+r_2)$, which can be written as
\begin{subequations}\begin{align}
\cos^2(\nicefrac{\theta}{2})|\sin(\chi_- \pm \nicefrac{\pi}{4})|&\leq 2^{-\nicefrac{1}{2}}\gamma^{-2}	\,,	\label{ineq:suba}\\
|a^2 - b^2 + 2a b| &\leq \gamma^{-2}	\label{ineq:subb}\,,
\end{align}\end{subequations}
where $\chi_{\pm}=\chi_1 \pm \chi_2$ and
\begin{align}
a &= \cos(\nicefrac{\chi_-}{2}) \cos(\nicefrac{\theta}{2})	\,,	\nonumber\\
b &= \cos(\nicefrac{\chi_+}{2}) \sin(\nicefrac{\theta}{2})	\,.
\end{align}
By adding multiples of $\nicefrac{\pi}{2}$ to $\chi_1$ and/or $\chi_2$ and changing the sign of $\chi_{\pm}$ (which will only permute the two inequalities in Eq.~\eqref{ineq:suba}), we can further restrict the parameters to the region
\begin{align}
\mathcal{V}=\left\{\theta,\chi_{\pm}|	\theta\in[0,\nicefrac{\pi}{3}],	\chi_-\in[0,\nicefrac{\pi}{4}],	\chi_+\in[0,\nicefrac{\pi}{2}]\right\}
\end{align}
and ignore the ``$-$'' inequality in Eq.~\eqref{ineq:suba} as it is not violated in $\mathcal{V}$. Therefore at a given noise level $\gamma$, the probability of the observers choosing measurements that result in nonlocal correlations is lower-bounded by
\begin{align}\label{eq:prob}
p(\gamma) &= |\mathcal{V}|^{-1}\int_\mathcal{V}d\theta d\chi_-d\chi_+ \, \sin\theta f(\theta,\chi_{\pm},\gamma)
\end{align}
where $|\mathcal{V}|$ is the volume of the region $\mathcal{V}$ and $f(\theta,\chi_{\pm},\gamma)=1$ if $(\theta,\chi_{\pm},\gamma)$ violate Eq.~\eqref{ineq:suba} or \eqref{ineq:subb} and 0 otherwise. 

In the absence of noise ($\gamma=1$), the only measurements that do not violate a Bell inequality are when the two parties' measurements are perfectly aligned, i.e., $\theta=0$ and $\chi_1=\chi_2$. As this is a set of measure zero, the parties always violate a Bell inequality.

\begin{figure}[t!]
\centering
\includegraphics[width=\linewidth]{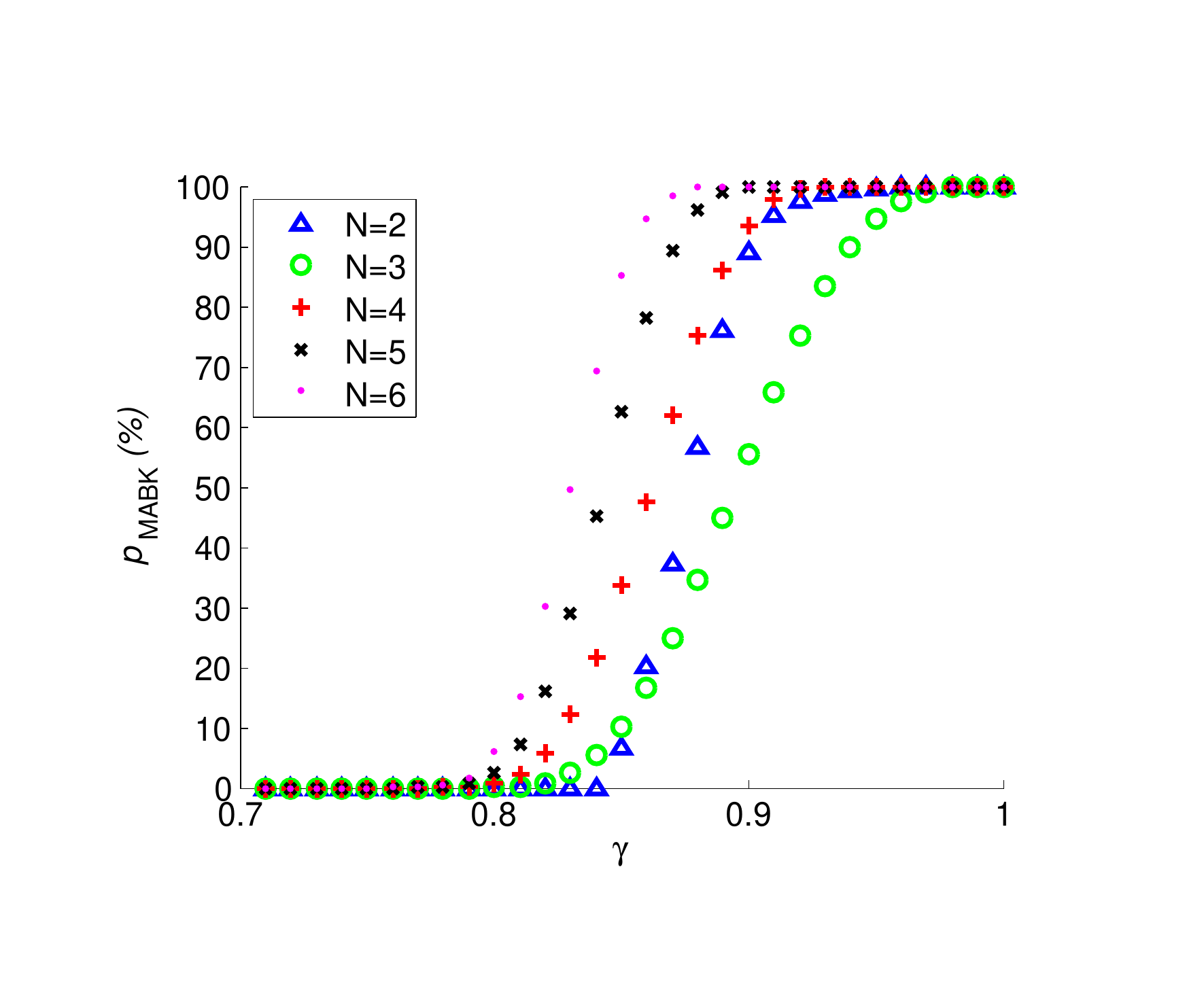}
\caption{\label{fig:mabk_noisy} (Color online) Plot of the probability of violating one of the MABK Bell inequalities, $p_{\rm MABK}$, for $N=2,\ldots,6$ parties as a function of the noise parameter $\gamma$ for local depolarizing noise.}
\end{figure}

As $\gamma$ decreases from 1, small perturbations from this perfect alignment also do not violate a Bell inequality. For mathematical convenience, we only consider noise levels $\gamma\geq \nicefrac{2}{\sqrt[4]{18}}\sim 0.97$ analytically. Numerical data for the full range of noise levels that allow violations of a Bell inequality (determined from Tsirelson's bound~\cite{cirelson}) is plotted in Fig.~\ref{fig:mabk_noisy}. 

For fixed $\gamma\in[\nicefrac{2}{\sqrt[4]{18}},1]$, the Bell inequalities in Eq.~\eqref{ineq:suba} and \eqref{ineq:subb} are violated for $(\theta,\chi_{\pm})\in\mathcal{V}$ when
\begin{align}
\chi_- >\sin^{-1}\left(2^{-\nicefrac{1}{2}}\gamma^{-2}\cos^{-2}(\nicefrac{\theta}{2})\right)-\nicefrac{\pi}{4} := L(\theta,\gamma)
\end{align}
or
\begin{align}\label{ineq:subd}
a - \sqrt{2a^2-\gamma^{-2}}	- 2^{-\nicefrac{1}{2}}\sin(\nicefrac{\theta}{2}) < 0 \,,
\end{align}
respectively, where we have used $a>b>2^{-\nicefrac{1}{2}}\sin(\nicefrac{\theta}{2})$ everywhere in $\mathcal{V}$. The left-hand side of Eq.~\eqref{ineq:subd} is convex in $\theta$ and nonincreasing in $\chi_-$ everywhere in $\mathcal{V}$. Therefore Eq.~\eqref{ineq:subd} will be satisfied for all $\theta\in[x(\gamma),\nicefrac{\pi}{3}]$ and $\chi_-\in [0, L(\theta,\gamma)]$ if it is satisfied for $\theta = x(\gamma)$, $\chi_- = L(x(\gamma),\gamma)$ and $\theta = \nicefrac{\pi}{3}$, $\chi_- = L(\nicefrac{\pi}{3},\gamma)$. 

Choosing $x(\gamma) = \cos^{-1}\gamma^{\nicefrac{1}{6}}$, these conditions are satisfied, so a Bell inequality is always violated unless $\theta \leq \cos^{-1}\gamma^{\nicefrac{1}{6}}$ and $\chi_-\leq L(\theta,\gamma)$. Therefore, for fixed $\gamma\in[\nicefrac{2}{\sqrt[4]{18}},1]$, the probability of not violating a Bell inequality is upper-bounded by
\begin{align}
1-p(\gamma) &\leq \nicefrac{8}{\pi}\int_0^{\cos^{-1}\gamma^{\nicefrac{1}{6}}}d\theta \sin\theta L_2(\theta,\gamma)	\nonumber\\
&\leq (1-\gamma^{\nicefrac{1}{6}})/4	\,.
\end{align}

For $\gamma\geq\nicefrac{2}{\sqrt[4]{18}}$, the probability of violating a Bell inequality is at least $99.8\%$, so the probability of violating a Bell inequality is remarkably robust against noise.

\textit{The multipartite case.}---We now consider $N$ parties who implement the same scheme using the $N$-partite GHZ state,
\begin{align}
\ket{\Psi^N_{\rm GHZ}} = 2^{-\nicefrac{1}{2}}(\ket{\vec{0}_N} + \ket{\vec{1}_N})\,,\label{eq:GHZ_state}
\end{align}
where $\ket{\vec{i}_N}$ denotes the state in which $N$ qubits are prepared in the state $\ket{i}$. For the $N$-partite GHZ state with local depolarizing noise and measurements parametrized as in Eq.~\eqref{eq:measurements}, the correlation functions are
\begin{align}\label{eq:GHZ_correlations}
E(\vec{s}) &= \text{Tr}\bigl(\rho(\lambda)\bigotimes_{j=1}^N \gamma\O^j_{s_j}\bigr)	\nonumber\\
&=\gamma^N\delta_N\prod_{j=1}^N ({\Omega}^j_{s_j})_z + \tfrac{\gamma^N}{2} {\rm Re}\prod_{j=1}^N [({\Omega}^j_{s_j})_x + i({\Omega}^j_{s_j})_y]	\,,
\end{align}
where $\delta_N \equiv 1-N$ (mod 2).

To obtain a numerical estimate of the probability of the $N$ parties violating a MABK Bell inequality without sharing a reference frame, we randomly sample $10^7$ sets of measurements according to the uniform Haar measure on the surface of the sphere and find the fraction of measurements that violate an MABK Bell inequality. The results are plotted in Fig.~\ref{fig:mabk_noisy} as a function of $\gamma$. 

For $N\neq 3$, all $10^7$ sets of measurements led to a violation of a MABK inequality, and so the numerical evidence suggests that the parties will always violate an MABK inequality and the robustness to noise increases with $N$. The exceptional case of $N=3$, for which the numerical probability of violating an MABK inequality in the absence of noise is ${\sim 99.99\%}$, occurs because the correlation functions are independent of the $z$-component of the measurements for odd $N$ (indicated by the $\delta_N$ term in Eq.~\eqref{eq:GHZ_correlations}).

\textit{Summary.}---We have proven that two parties who share a maximally entangled bipartite state will always violate a Bell inequality by choosing random measurements from a triad of measurements (corresponding to the $x$, $y$ and $z$ directions on their local Bloch sphere). We have also provided numerical evidence that $N$ parties who share a maximally entangled state will always violate a Bell inequality (unless $N=3$) with this measurement choice. Moreover, this scheme is robust against local depolarizing noise in that small levels of noise will only slightly decrease the probability of violating a Bell inequality.

We note that local depolarizing noise models a variety of relevant experimental noise sources and imperfections. For example, local depolarizing noise can be used to model imperfect detectors (i.e., detectors that only detect a fraction $\gamma$ of events) or the non-ideal preparation of a resource state through such processes as spontaneous parametric down conversion. Local depolarizing noise also provides a worst-case bound for other noise models. Finally, the singlet state with local depolarizing noise is equivalent to a mixed Werner state, so our results can also be interpreted as giving a probability of violating a Bell inequality for Werner states.

\begin{acknowledgments}
JJW thanks Yeong-Cherng Liang for communicating the numerical observation in the bipartite scenario, which inspired the results presented here and was also considered independently in Ref.~\cite{Shadbolt2011}. This research is supported by the Australian Research Council.
\end{acknowledgments}

\end{document}